\documentclass[twocolumn,showpacs,showkeys,preprintnumbers,amsmath,amssymb]{revtex4}
\usepackage [dvips]{graphicx}

\usepackage [english]{babel}

\begin{document}

\title{Effects of broadening and electron overheating
in tunnel structures based on metallic clusters}

\author{A. V. Babich and V. V. Pogosov\footnote{Corresponding author: E-mail:
 vpogosov@zstu.edu.ua }}

\address{Department of Micro- and Nanoelectronics, Zaporozhye National
Technical University, Zhukovsky Str. 64, Zaporozhye 69063, Ukraine}

\begin{abstract}
We study the influence of energy levels broadening and electron
subsystem overheating in island electrode (cluster) on
current-voltage characteristics of three-electrode structure. A
calculation scheme for broadening effect in one-dimensional case is
suggested. Estimation of broadening is performed for electron levels
in disc-like and spherical gold clusters. Within the two-temperature
model of metallic cluster and by using a size dependence of the
Debye frequency the effective electron temperature as a function of
bias voltage is found approximately. We suggest that the effects of
broadening and electron overheating are responsible for the strong
smoothing of current-voltage curves, which is observed
experimentally at low temperatures in structures based on clusters
consisting of accountable number of atoms.
\end{abstract}


\maketitle

\section{Introduction}

The nanodispersed metallic systems are prospective objects of
nanotechnology. Therefore understanding of their physical properties
is of scientific and hopefully of practical interest.

Transport of electrical charge across a nanoscale tunnel junction is
accompanied by many effects, such as the Coulomb blockade of the
average current, transfer of energy between electrons and ions, and
consequently heating of the junction. In nanometer scale devices
electron transport can occur through well-resolved quantum states.
If the temperature is increased, the Coulomb and quantum staircases
of current are gradually smeared out by thermal fluctuations (see,
for example \cite{Gub-98,Ralph}).

Simple tunnel construction can be schematically represented by the
distinctive ``sandwich'' \cite{Wang,Hou,Ohgi,Ohgi-2003}: Au(111)
thick film / dielectric nanofilm / isolated Au cluster / vacuum gap
/ polycrystalline Au (a tip of STM). The monoatomic disc-shaped
\cite{Wang,Hou} or spherical-like \cite{Ohgi,Ohgi-2003} gold
clusters are self-organized on the dielectric layer.

Some of the experimental features of the $I-V$ curves were
investigated in Ref. \cite {FTT-2006} including the current gap in
the low temperature limit \cite {LetJTF-2007}. However, the fact of
smoothing of staircases for granule-molecule at low temperatures is
still not understood \cite {Gub-98,Ralph,Wang,Hou,Ohgi,Ohgi-2003}.
Such a smoothing is typical for molecular transistors. Moreover, the
observed current gap decreases significantly as temperature
increases from 5 K to 300 K in structure based on disk-shaped
cluster with the radius $\sim$2 nm \cite{Wang}. However, for
spherical granules, of radius $\sim$1 nm, a similar dependence of
current gap cannot be traced back by a comparison of gaps at $T=30$
\cite{Ohgi} and 300 K \cite{Ohgi-2003}.

Such ``anomaly'' of the temperature dependence in the regime of the
Coulomb blockade and strong quantization can hardly be explained
within the concept of a quasi-equilibrium electronic gas and
resonance tunneling through the stationary electronic states.

The aim of this paper is to analyze two mechanisms: (i) the
broadening of electronic levels due a tunnel effect, (ii) electronic
gas heating in the isolated metallic clusters of disc-like and
spherical shape in presence of bias voltage. Absence of clear steps
of the Coulomb and quantum staircases on the experimental
current-voltage curves of single-electronic devices at the low
temperatures is explained by these effects.

In the typical electronic circuit the current flow leads to the
non-equilibrium regime of interaction between electron and the
phonon (ion) subsystems. Dynamics of a relaxation of non-equilibrium
electrons was examined in metals \cite{ginz,kag,zl}, continuum films
\cite{Yan,Gloskovskii}, nanowires
\cite{Kopidakis,Qu,Vega,Agosta,Galperin}, particle films
\cite{Gloskovskii,BLT,Fed,Bil,Singh,Ovadia}, free clusters
\cite{Kres,Pushpa}, including the regime under the action of the
piko- and femtosecond laser pulses
\cite{Gloskovskii,Rethfeld,anis-2004,Ji}. Very few papers deal with
the direct experiments with the free metal clusters within
ultrashort pulse duration (see \cite{Gantefor,Maier} and references
therein). A power injected by a laser in metals and clusters within
the time of pulse allows one to trace directly a kinetics of the
relaxation between electrons and lattice.

Predicted earlier size dependence of the Debye temperature \cite
{Co} is experimentally confirmed in Ref. \cite {Balerna} and then it
is precised  by temperature dependence in Ref. \cite {Gu}.
Suppression of the electron-phonon interactions in granules is a
result of deformations of a phonon spectrum in these systems. This
interaction can to be suppressed so that electron-electron
interaction appears as a basic mechanism of dissipation affecting a
particle energy. This leads to the overheating of electronic
subsystem which can be described by Fermi statistics with some
effective (raised) temperature while the ionic subsystem temperature
varies only slightly.

It is supposed that the relaxation of the non-equilibrium electrons
in small metal particles, films \cite{BLT,Fed,Bil,Singh} and wires
\cite{Zhang,Urban} occurs owing to  excitation of the Rayleigh waves
or surface acoustic phonons. However, the obtained expressions in
the cited works contain no asymptotic transition to infinite
systems.

According to the Weyl`s theorem (see Ref. \cite {Mar}), it is
possible to separate the bulk and surface acoustical phonons only
for the large metal sample. For small-sized samples the modes are
mixed, a sound velocity becomes indefinite and, as a rule, in
practice it is used as a fitting parameter. On the other hand,
measurements of an electron-ion power exchange in the free clusters
Na$_{16-250}^{+}$ \cite{Maier} have demonstrated that, for
reasonable estimations, it is quite possible to use the conception
of bulk phonons, but with the account of the size dependence of the
Debye frequency. Such an approach for the metallic nanoclusters,
films and wires can be considered as an extrapolation.

\section{Broadening of levels}

The scattering matrix relates the initial state and the final state
for an interaction of particles. The free electronic states of the
cluster do not decay so that they are stationary. For free clusters
the poles of the scattering matrix $S(k)$, located on the real axis
at the plane of complex values of waving numbers $k$, correspond to
the stationary states.

If a cluster is placed between electrodes, its electronic states
become quasi-stationary. Broadening occurs due to a tunneling effect
by analogy with the formation of energy bands in a crystal. The
broadening increases with the increase of the bias voltage applied
between electrodes. Both tunnel barriers are three-dimensional.
Therefore, the problem of calculation of broadening, in general
case, is far from being trivial. An analytical solution of resonance
tunneling problem is possible only for one-dimensional geometry and
rectangular barriers (see \cite{288}).

According to the indeterminacy principle broadening effects are
related to the finite life time. Energies of the quasi-stationary
states become complex and their imaginary parts describe levels
broadening. The poles of the $S-$matrix corresponding to these
states are located in the lower half of the complex plane wave
number $k$. The state with well defined energy is accordingly
replaced by the Lorentz distribution with the scale parameter which
specifies the half-width at half-maximum $\gamma_{p}$:
\begin{equation}
L_{p}(\varepsilon)=\frac{1}{2\pi}\frac{\gamma_{p}}
{(\varepsilon-\varepsilon_{p})^{2}+\gamma_{p}^{2}/4}. \label{BP-1}
\end{equation}
Here an index $p$ denotes  the set of quantum numbers (except for a
spin) which correspond to the single-electron state with energy
$\varepsilon_{p}$ (Fig. 1). $L_{p}(\varepsilon)\rightarrow
\delta(\varepsilon-\varepsilon_{p})$ as $\gamma_{p}\rightarrow 0$
where $\delta(x)$ is the Dirac $\delta-$function.

Taking into account the broadening function, the electron density of
states can be expressed as
\begin  {equation}
\overline{\rho}(\varepsilon)=2\sum_{p}L_{p}(\varepsilon),
\label{BP-1}
\end{equation}
where the factor 2 takes into account a spin degeneracy.
\begin{figure}[!t!b!p]
\centering
\includegraphics [width=.49\textwidth] {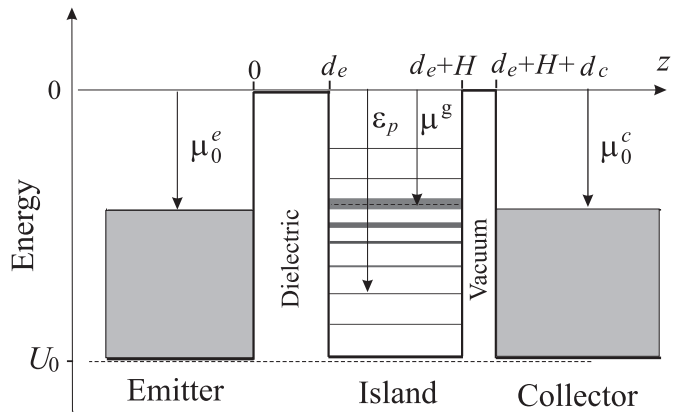}
\caption{The energy diagram for tunnel structure based on the
non-magic granule before application of voltage.}
\label{BP-2009-Fig1}
\end{figure}

Current flowing through a quantum granule (with limitation on its
Coulomb instability \cite{FTT-2006}) is determined by the equality
condition between the emitter and collector currents ($I^{\rm
e}=I^{\rm c}\equiv I$) or
\begin{eqnarray}
-e\sum_{n_{\rm min}}^{n_{\rm max}} P_{ n}\left(\overrightarrow{w _{
n}^{\rm  e}}-\overleftarrow{w _{ n}^{\rm  e}}\right)= -
e\sum_{n_{\rm min}}^{n_{\rm max}} P_{ n}\left(\overrightarrow{w _{
n}^{\rm c}}-\overleftarrow{w _{ n}^{\rm  c}}\right). \label{set-60}
\end{eqnarray}
The probability $P_{n}$ of finding of $n$  ``surplus'' ($n>0$) or
failing ($n<0$) electrons at central electrode is defined by the
master equation in the stationary limit. In reality, one calculates
the reduced current $\widetilde{I}\equiv I/(eP_{0}\Gamma^{\rm e})$
where $\Gamma^{\rm e,c}$ are tunnel rates, ($-e$) is the electron
charge. In order to find $P_{n\neq 0}/P_{0}$ the recurrent relation
is used:
\begin{equation}
P_{n+1}=P_{ n}\,\frac{w_{n}^{\rm in}}{w_{n+1}^{\rm out}}
\label{set-58}
\end{equation}
where $w_{n}^{\rm in}=\overrightarrow{w_{n}^{\rm
e}}+\overleftarrow{w_{ n}^{\rm c}}$ and $w_{ n}^{\rm
out}=\overleftarrow{w_{ n}^{\rm e}}+\overrightarrow{w_{n}^{\rm c}}$
are the total  electron streams from/to leads into/out the cluster,
and $\overleftarrow{\overrightarrow{w _{ n}^{\rm e,c}}}$ are the
partial tunneling streams, accordingly. Here the upper/under arrows
and indexes ``{\rm e,c}'' denote the emitter-granule and
collector-granule and back transitions in accordance with arrows
direction.

Taking into account the broadening of levels for $V>0$, we have
\begin{eqnarray}
\overrightarrow{w_{n}^{\rm e}}&=&\frac{1}{\pi}\Gamma^{\rm e}\sum_{
p}\int\limits_{U_{0}+\overrightarrow{U^{e}}}^{+\infty}\,
\frac{\gamma(\overrightarrow{\varepsilon^{\rm
e}})}{(\varepsilon'-\overrightarrow{\varepsilon^{\rm
e}})^{2}+(\gamma(\overrightarrow{\varepsilon^{\rm
e}})/2)^{2}}\times \nonumber\\
& &f(\varepsilon'-\mu_{V}^{\rm
e};\,T)\,[1-f(\varepsilon'-\overrightarrow{\mu_{C}^{\rm
e}};\,T_{e})]\,d\varepsilon', \label{set-55}
\end{eqnarray}
\begin{eqnarray}
\overleftarrow{w_{ n}^{\rm c}}&=&\frac{1}{\pi}\Gamma^{\rm c}\sum_{
p}\int\limits_{U_{0}+\overleftarrow{U^{c}}}^{+\infty}\,
\frac{\gamma(\overleftarrow{\varepsilon^{\rm
c}})}{(\varepsilon'-\overleftarrow{\varepsilon^{\rm
c}})^{2}+(\gamma(\overleftarrow{\varepsilon^{\rm
c}})/2)^{2}}\times \nonumber\\
& &f(\varepsilon'-\mu_{V}^{\rm c};\,T)\, [1-f(\varepsilon'-
\overleftarrow{\mu_{C}^{\rm c}};\,T_{e})] \,d\varepsilon',
\label{set-55a}
\end{eqnarray}
\begin{eqnarray}
\overleftarrow{w_{ n}^{\rm e}}&=&\frac{1}{\pi}\Gamma^{\rm
e}\sum_{p}\int\limits_{U_{0}+\overleftarrow{U^{e}}}^{+\infty}\,
\frac{\gamma(\overleftarrow{\varepsilon^{\rm
e}})}{(\varepsilon'-\overleftarrow{\varepsilon^{\rm
e}})^{2}+(\gamma(\overleftarrow{\varepsilon^{\rm
e}})/2)^{2}}\times \nonumber\\
& &[1-f(\varepsilon'-\mu_{V}^{\rm
e};\,T)]\,f(\varepsilon'-\overleftarrow{\mu_{C}^{\rm
e}};\,T_{e})\,d\varepsilon', \label{set-55b}
\end{eqnarray}
\begin{eqnarray}
\overrightarrow{w_{ n}^{\rm c}}&=&\frac{1}{\pi}\Gamma^{\rm c}\sum_{
p}\int\limits_{U_{0}+\overrightarrow{U^{c}}}^{+\infty}\,
\frac{\gamma(\overrightarrow{\varepsilon^{\rm
c}})}{(\varepsilon'-\overrightarrow{\varepsilon^{\rm
c}})^{2}+(\gamma(\overrightarrow{\varepsilon^{\rm
c}})/2)^{2}}\times \nonumber\\
& &[1-f(\varepsilon'-\mu_{V}^{\rm c};\,T)]\,
f(\varepsilon'-\overrightarrow{\mu_{C}^{\rm c}};\,T_{e})
\,d\varepsilon' \label{set-55c}
\end{eqnarray}
where $f(\varepsilon -\mu;\,T)=\left\{1+\exp[(\varepsilon
-\mu)/k_{\rm B}T]\right\}^{-1}$ is the Fermi-Dirac distribution.

Because of the applied voltage and charging of a granule
\cite{FTT-2006}, the spectrums  and the chemical potentials are
shifted:
$$
\overrightarrow{\overleftarrow{\varepsilon^{\rm e}}}= \varepsilon_{
p}+\widetilde{E}_{\rm C}(n \pm 1/2)- e\eta^{+} V,
$$
$$
\overrightarrow{\overleftarrow{\varepsilon^{\rm c}}}= \varepsilon_{
p}+\widetilde{E}_{\rm C}(n \mp 1/2)+e(1-\eta^{+}) V,
$$
$$
\overleftarrow{\overrightarrow{U^{\rm e}}}=-e\delta
\phi+\widetilde{E}_{\rm C}(n \mp 1/2)-å\eta^{+} V,
$$
$$
\overleftarrow{\overrightarrow{U^{\rm c}}}=-e\delta
\phi+\widetilde{E}_{\rm C}(n\pm 1/2)+e(1-\eta^{+}) V,
$$
$$
-\mu_{V}^{\rm e}\equiv W_{0}^{\rm e},\quad
\overleftarrow{\overrightarrow{\mu_{C}^{\rm e,c}}}=\mu^{\rm
g}+\overleftarrow{\overrightarrow{U^{\rm e,c}}}, \quad \mu_{V}^{\rm
c}= \mu_{0}^{\rm c}-eV.
$$
Here the upper/under arrows at the left correspond to the following
signs at the right. $\varepsilon_{p}$ is  electron spectrum in a
cluster in absence of both the voltage and charging,  $W_{0}^{\rm
e}\equiv -\mu_{0}^{\rm{e}}$ is a work function of semi-infinity
metal, $\mu^{\rm {g}}$ is a  electron chemical potential of granule,
$\delta \phi=(\mu^{ \rm {g}}-\mu_{0}^{\rm{e,c}})$ is a contact
potential difference between cluster and electrodes.

For $V>0$ the fraction of voltage reads
\begin{equation}
\eta^{+}=\frac{d_{\rm e}+\epsilon L/2}{\epsilon(d_{\rm c}+L)+d_{\rm
e}} \label{frac}
\end{equation}
where $L\equiv 2R$ and $H$ for a sphere and disk, accordingly,
$\epsilon$ is a dielectric constant of film which covers the left
electrode. $\eta^{+} V$ is the potential in a coordinate
$z=d_{e}+L/2$ in the case of absence of cluster (it is assumed that
the electric field in the cluster is screened completely). For $V<0$
the voltage fraction $\eta^{-}$ equals $1-\eta^{+}$.

As an approximation, the profile of the one-electron effective
potential in the cluster can be represented as a potential well of
the depth $U_{0}<0$. The three-dimensional Schr\"{o}dinger equation
for a quantum box can be separated into one-dimensional equations.
The spectrum of wave numbers in a spherical and cylindrical
potential wells are determined from the continuity condition of a
logarithmic derivative of the wave function on the boundaries.

Neglecting the area near cylinder edges, the energy spectrum in
metal nanodisk is found by a simple way as follows
\begin{equation}
\varepsilon_{p}=U_{0}+\frac{\hbar^{2}}{2m_{e}}(k_{n_{z}}^{2}+k_{\perp}^{2}
) \label{EP}
\end{equation}
where $U_{0}<0 $ is the position of conductivity band of a
semi-infinite metal \cite{PHR-2005}, $k_{\perp}$ is a solution of
wave equation for radial direction. Quantization of the wave number
$k_{n_{z}}$ along the cylinder axis is determined by the solution of
the equation:
\begin{equation}
k_{n_{z}}H =n_{z}\pi -2\arcsin (k_{n_{z}}/k_{0}) \label{EZ}
\end{equation}
where $n_{z}$ is the integer number, $\hbar k_{0}\equiv
\sqrt{2m_{e}|U_{0}|}$. Since the tunneling takes place mainly in
$z-$direction, ``partial'' broadening of  $k_{n_{z}}$ spectrum
corresponds to \emph{general} spectrum $\varepsilon_{p}$.

In order to calculate the electron levels broadening in quantum
metal disk, let us consider the decay of cluster's states due to the
tunneling. For simplest potential profile which corresponds to Fig.
1, we use the solution of the Schr\"{o}dinger equation in the form:
\begin{multline}
\overrightarrow{\psi(z)}= \\
\left\{
\begin{array}{ll}
   e^{ik_{{n_{z}}}z}+B_{1}e^{-ik_{{n_{z}}}z}, \,\, & z < 0, \\
   A_{1}e^{\kappa_{{n_{z}}}z}+B_{2}e^{-\kappa_{{n_{z}}}z}, \,\, & 0 < z < d_{e}, \\
   A_{2}e^{ik_{{n_{z}}}z}+B_{3}e^{-ik_{{n_{z}}}z}, \,\, & d_{e}< z < d_{e}+H, \\
   A_{3}e^{\kappa_{{n_{z}}}z}+B_{4}e^{-\kappa_{{n_{z}}}z}, \,\, & d_{e}+H < z < d_{e}+H+d_{c}, \\
   A_{4}e^{ik_{{n_{z}}}z}, \,\, & z > d_{e}+H+d_{c} \\
\end{array}
\right. \label{U002}
\end{multline}
for the  electrons stream falling from the left to the right and
\begin{multline}
\overleftarrow{\psi(z)}= \\
\left\{
\begin{array}{ll}
   e^{-ik_{{n_{z}}}z}+B_{5}e^{ik_{{n_{z}}}z}, \,\, & z > d_{e}+H+d_{c}, \\
   A_{5}e^{-\kappa_{{n_{z}}}z}+B_{6}e^{\kappa_{{n_{z}}}z}, \,\, & d_{e}+H < z < d_{e}+H+d_{c}, \\
   A_{6}e^{-ik_{{n_{z}}}z}+B_{7}e^{ik_{{n_{z}}}z}, \,\, & d_{e}< z < d_{e}+H, \\
   A_{7}e^{-\kappa_{{n_{z}}}z}+B_{8}e^{\kappa_{{n_{z}}}z}, \,\, & 0 < z < d_{e}, \\
   A_{8}e^{-ik_{{n_{z}}}z}, \,\, & z < 0 \\
\end{array}
\right. \label{U003}
\end{multline}
for the stream  falling from right to left, accordingly. According
to (\ref{U002}) and (\ref{U003}), $\hbar \kappa_{n_{z}}$ is equal to
$\sqrt{2m_{e}|U_{0}|-\hbar^{2}k_{{n_{z}}}^{2}}$.

Using the continuity condition of the wave functions on the
boundaries $z=0,\,d_{e},\,d_{e}+H$ and $d_{e}+H+d_{c}$ we obtain the
system of equations for the determination of the coefficients $A$
and $B$ which we then solve numerically by the LU-expansion method.

A total wave function can be written using the $S-$matrix as
$$
\psi(z)\sim \{\overleftarrow{\psi(z)}-S\overrightarrow{\psi(z)}\}.
$$
For any coordinates inside the electron reservoirs (left and right
electrodes), $z=z^* \leq 0$ or $z^*\geq d_{e}+H+d_{c}$ (Fig. 1), one
can calculate the matrix
\begin{equation}
S=(\overleftarrow{\psi}/\overrightarrow{\psi})|_{z=z^{*}}.
\label{dissip}
\end{equation}

By the Muller`s method we calculate the pole of $S-$matrix at the
lower half-plane of the complex wave numbers $k$, in the vicinity of
point $k_{{n_{z}}}$. The imaginary part of the energy
$\hbar^{2}k_{{n_{z}}}^{2}/2m_{e}$ gives the energy broadening.

It is easy to generalize a method on $V \neq 0$ regime.  In this
case underbarrier wave functions will be expressed through the Airy
functions.

For estimation of the broadening in a spherical cluster, it is
possible to use the solution of the well-known problem for open dot
-- spherically symmetric potential in depth $U_{0}$, of radius  $R$
and barrier thickness $d_{c}$. We define broadening by analogy with
the book \cite{288} as
\begin{equation}
\gamma_{p}\approx 8 e^{-2\kappa {p}
d_{c}}\frac{\hbar^{2}k_{p}^{3}\kappa_{p}^{3}}{m_{e}k_{0}^{4}(1+\kappa_{p}d_{c})}.
\label{gamma}
\end{equation}

\section{Balance Equation}

The two-temperature model describes a system of electrons and ions,
which is out of equilibrium between electronic and ionic subsystems.
For a metallic sample, this condition can be fulfilled, provided one
applies an electric field.

In a two-temperature model a balance equation in a cluster in
presence of voltage has the simplest form
$$
\Omega\frac{\partial (c_{e}T_{e})}{\partial t}= P(T_{e},T_{i})
-Q(T_{e},T_{i}),
$$
\begin{equation}
\Omega\frac{\partial (c_{i}T_{i})}{\partial t}= Q(T_{e},T_{i})
\label{Bal-1}
\end{equation}
where $c_{e,i}$ is specific heat capacity of electronic and ionic
subsystems (with temperatures $T_{e}$ and $T_{i}$, respectively) of
cluster with the volume $\Omega$, $P$ is a input power, $Q$ is the
exchange energy between electrons and ions per second.

Since the specific heat of the electronic subsystem is much smaller
than that of phonons, the electron-electron and the phonon-phonon
processes are much faster than the electron-phonon processes, i.e.
the characteristic relaxation time for the electron subsystem
temperature is much shorter than that for the phonon subsystem. The
result is that when injecting power into the metal cluster, the
electron temperature grows very rapidly until the energy flux from
electrons to phonons becomes equal to the absorbed power so that the
local equilibrium in the electron subsystem is achieved
($dT_{e}/dt=0$),
\begin{equation}
P(T_{e},T_{i}) -Q(T_{e},T_{i})=0. \label{Bal-2}
\end{equation}

For $Q$  we use the result of Ref. \cite{kag}, obtained for the case
of a massive metal, on the basis of kinetic equation:
\begin{multline}
Q(T_{e},T_{i})=\Omega\frac{2}{(2\pi)^{3}}\frac{m_{e}^{2}U^{2}_{e-ph}k_{\rm
B}^{5}T_{\rm D0}^{5}}{\hbar^{7}\rho s^{4}}\times\\
\left\{\left(\frac{T_{e}}{T_{\rm D0}}\right)^{5}\int\limits^{T_{\rm
D0}/T_{e}}_{0}\frac{x^{4}dx}{e^{x}-1}-\left(\frac{T_{i}}{T_{\rm
D0}}\right)^{5}\int\limits^{T_{\rm
D0}/T_{i}}_{0}\frac{x^{4}dx}{e^{x}-1}\right\}. \label{Bal-4}
\end{multline}
Here $U_{e-ph}$ is the electron-phonon interaction constant, $T_{\rm
D0}$ is the Debye temperature in a massive metal, $\rho$ is the
density of Au, and $s$ is the  ``average'' sound speed \cite{Gu}. In
literature, it is accepted to use the following expression for
$T_{e},T_{i}\gg T_{\rm D0}$ in the Eq. (\ref{Bal-4})
$$
Q(T_{e},T_{i})=\Omega\alpha(T_{e}-T_{i}).
$$

For low-dimensional object of a volume $\Omega$ and surface area $S$
the size dependence of the Debye temperature in quasi-classical
approximation is given by \cite{Co}:
\begin{equation}
T_{\rm D}=T_{\rm
D0}\frac{1+\pi\xi/8}{1+\pi\xi/4+(\xi/3)^{2}},\,\,\,\,\ \xi=
\frac{1}{k_{\rm WS}}\frac{S}{\Omega} \label{Co-Ded}
\end{equation}
where $k_{\rm WS}=(6\pi^{2}/\upsilon)^{1/3}$ is the maximum wave
number in a massive metal, $\upsilon=4\pi r_{0}^{3}/3$, and $ r_{0}$
is the atom density parameter ($ r_{0}=3\,a_{0}$ for Au). Reasonable
accuracy of the expression (\ref{Co-Ded}) was demonstrated in
experiment \cite{Balerna} where x-ray scattering was studied on gold
clusters with diameters ranging from 1.5 to 4.3 nm.

The feeding power can be counted up in the form
$P^{\pm}=I^{\pm}\eta^{\pm}V^{\pm}$ using the experimental $I(V)$
dependence. After that, the expression (\ref{Bal-4}), in which a
replacement $T_{\rm D0}\rightarrow T_{\rm D}$ (\ref{Co-Ded}) is
performed, is substituted in (\ref{Bal-2}). Under the assumption of
the equality between the temperatures of the ionic subsystem $T_{i}$
(constant throughout the tunnel structure) and thermostat, and from
the solution of (\ref{Bal-2}), one can find an electronic
temperature $T_{e}$, which characterizes the Fermi distribution,
(see (\ref{set-55}) and (\ref{set-55c})).

\section{Results and discussion}

We consider Au disks of monoatomic thickness whose radii vary in the
range $2R \simeq \{1,\,8.5\}$ nm and which contain $\simeq \{14,\,
10^{3}\}$ atoms. Similarly, the spherical clusters with $2R \simeq
\{1.4,\, 2.8\}$ nm contain $\simeq \{100,\,600\}$ atoms. (In Refs.
\cite{Ohgi,Ohgi-2003} cluster sizes are given in terms of monolayer
numbers; therefore, we used normalized curve from Fig. 1 of Ref.
\cite{Ohgi-2001}) in order to express these sizes in terms of
nanometers.)

The characteristic Coulomb energy of charging is $e^{2}/C$ where $C$
is self-capacitance of a single granule in vacuum. The calculations
of Ref. \cite{FTT-2006} demonstrated that these $C$ values are too
small for the width of the current gap to be explained. Therefore we
determine the characteristic energy of clusters charging as
$\widetilde{E}_{\rm C}=e^{2}/C_{\rm eff}$. Effective capacitance
$C_{\rm eff}=(R+\delta)$ is used in order to explain experimental
results for spherical clusters. The additional small quantity
$\delta$ is caused by an increase of radius of the charging electron
``cloud''. For gold $\delta$ is equal approximately to $ 1.8\,a_{0}$
\cite{PHR-2005}. The most obvious example is the case of a disc,
since almost half of the disc surface contacts to the dielectric
film with  $\epsilon=3$. In this case $C_{\rm eff}$ is estimated as
a capacitance of the spheroid with minor axis of length $H$. A major
axis $a$ is obtained from a condition $\pi R^{2}H=4\pi
a(H/2)^{2}/3$. Thus, we  have
$$
C_{\rm
eff}=\frac{1+\epsilon}{2}\frac{\sqrt{a^{2}-(H/2)^{2}}}{\arccos(H/2a)}.
$$
We note that the value of the capacitance is sensitive to the shape
of the granule surface so that even small deviation from the
spherical shape can change significantly the capacitance.

In this work calculations  are performed for structures based on
clusters, for which $I-V$ curves were measured at different
temperatures, namely, for a disk with a diameter $2R = (4\pm 0.5)$
nm and thickness $H \approx 0.3$ nm \cite{Wang} and for spheres with
$2R =(2\pm 0.35)$ nm \cite{Ohgi,Ohgi-2003}. Because of the
uncertainly of sizes and number of atoms, we used the jellium model
and found that disc and sphere contain 240 and 248 atoms,
accordingly. Then $\widetilde{E}_{\rm C}=$ 0.44 and 1.31 eV for the
disc and sphere, respectively.  In spite of the fact that volumes of
these two clusters are nearly the same, a difference between their
shapes produces a significant mismatch in $\widetilde{E}_{\rm C}$.

Clusters under consideration are non-magic. The Fermi level
$\mu^{g}$ and levels of lowest unoccupied $\varepsilon^{\rm LU}$ and
occupied $\varepsilon^{\rm HO}$ electron states in the clusters are
in line. Spectrums were calculated and reported in our earlier works
\cite {FTT-2006,LetJTF-2007}.

Taking into consideration the conditions of experiments
\cite{Wang,Hou,Ohgi,Ohgi-2003} and the symmetry of measured $I-V$
curves, the following numbers have been chosen as input parameters
in our calculations: $d_{e}=10$ {\AA}, $d_{c}=2$ {\AA} (Fig. 1), and
$\beta\equiv\Gamma^{\rm e}/\Gamma^{\rm c}=2$ and 1/2 for structures
based on a disk and sphere, accordingly.

Calculated $I-V$ curves for different magnitudes of a parameter
$\beta$ were analyzed in Ref. \cite{FTT-2006} where the effects of
broadening and overheating were neglected. As follows from the
expression
$$ \Delta
V_{g}=\frac{\widetilde{E}_{\rm C}}{2e}
\Big(\frac{1}{2-\eta^{+}}+\frac{1}{2-\eta^{-}}\Big),
$$,
the current gap is independent on $\beta$. However, the current
jumps are very sensitive to the value of $\beta$ which, in its turn,
has no influence on threshold voltages. With the growth of $\beta$,
the steepness of $I-V$ curves parts which correspond to $V>0$/$V<0$,
decreases/increases, respectively. For granular films a theory of
Ref. \cite{Imamura} gives a similar result, however, measurements of
Ref. \cite{Imamura} demonstrate the influence of tunneling
resistances (in other words, of parameter $\beta$) on the current
gap width.

We perform calculations for gold clusters with the electron-phonon
interaction constant $U_{e-ph}=1$ eV \cite{Singh}, the Debye
temperature $T_{\rm D0}=150$ K, the density of Au $\rho=19.3\times
10^{3}$ kg/m$^{3}$, and the  ``average'' sound speed $s=1500$ m/s
\cite{Gu}.

The size dependences of the Debye temperature $T_{\rm D}(R)$ were
analyzed taking into account Eq. (\ref{Co-Ded}). The actual forms
$T_{\rm D}(R)$ in a wide range of sizes are plotted in Fig. 2.
Different asymptotic behaviors of these two curves is due to the
fact that in Eq. (\ref{Co-Ded}) at $R \rightarrow \infty$, one has
$S/\Omega \rightarrow 0$ for spheres and $2/H$ for discs.
\begin{figure}[!t!b!p]
\centering
\includegraphics [width=.45\textwidth] {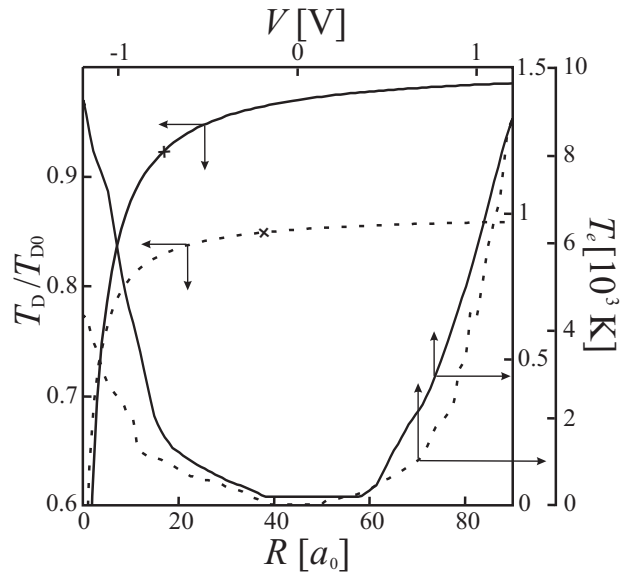}
\caption{Size dependences of the Debye temperature $T_{\rm D}(R)$ in
spheres (continuous line)  and discs (dotted line). Values of
$T_{\rm D}(R)$ are depicted ($\times$) at the curves, and these
values are then used in calculations of electron heating in the
sphere with $R=1$ nm and disc with $R=2$ nm. For these two clusters,
voltage dependences of electronic kinetic temperature $T_{e}(V)$ are
presented too. } \label{BP-2009-Fig2}
\end{figure}
\begin{figure}[!t!b!p]
\centering
\includegraphics [width=.45\textwidth] {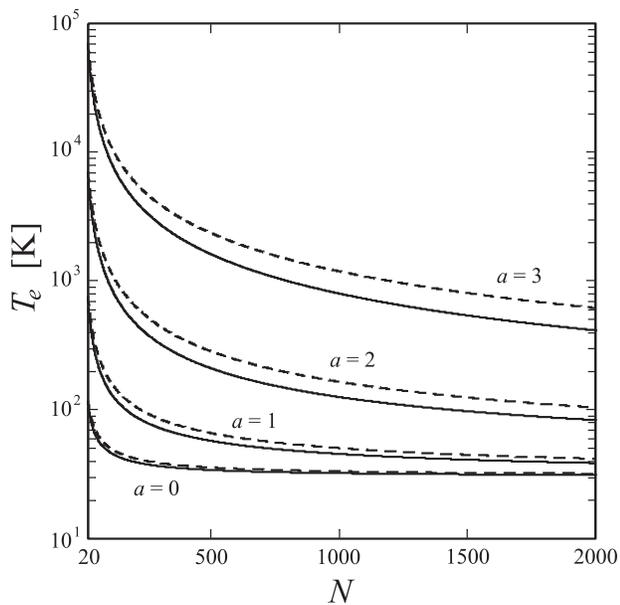}
\caption{Size dependences of the electronic kinetic temperature
$T_{e}(N)$ in spheres (continuous line)  and discs (dotted line) for
different values of injected power $P=10^{a}P_{0}$, $P_{0}=10^{-12}$
W. $N$ is the number of atoms.} \label{BP-2009-Fig3}
\end{figure}

The feeding power leads to the overheating of the electron
subsystem. With the increase of the bias voltage $V$ the number of
electrons relaxing in the granule increases significantly. Among
them are all the electrons with energies in the interval $e\eta V$
below the Fermi level of the granule, since the flow of tunneling
electrons increases from below lying levels, thereby, involving
large number of conductivity electrons to the relaxation process.
The granule does not fragmentize during the significant overheating
of the electron subsystem, because the $I-V$ curves are reproduced
during the cyclic changes of the bias voltage
\cite{Wang,Hou,Ohgi,Ohgi-2003}.

The dependences $T_{e}(V)$  for two structures based on sphere  and
disk (temperature of ions 5 K and 30 K, accordingly) are also
plotted
 in Fig. 2. One can observe strong dependence of electronic
kinetic temperature on voltage. Heating of electrons in a disk is
much more intensive than in a sphere: the corresponding temperature
is almost one order of magnitude higher and it achieves thousands of
Kelvins. It is of interest to note that the tunneling current in 1
pA only is provided by $\sim 10^{6}$ of electrons per second that is
a significant number for the granule, which contains hundreds of
conductivity electrons.

One of the conclusions of recent article \cite{Gloskovskii} is the
fact of the increase of the kinetic electron temperature with the
decrease of cluster size for $P=$const. The results of our
calculations, which are presented in Fig. 3, confirm this
conclusion.

The calculated $I-V$ characteristics for structures, based on
spherical and disk clusters, are plotted in the Figs. 4 and 5. At
low temperatures, the calculated values of gap width $\Delta V_{g}$
are in accordance with the experimental data for a structure on
disc-like cluster (Fig. 5). A difference (approximately in 1.5
times) for a structure on a spherical cluster (Fig. 4) can be
possibly attributed to
the fact that we have neglected the mutual capacities effect. 
\begin{figure}[!t!b!p]
\centering
\includegraphics [width=.45\textwidth] {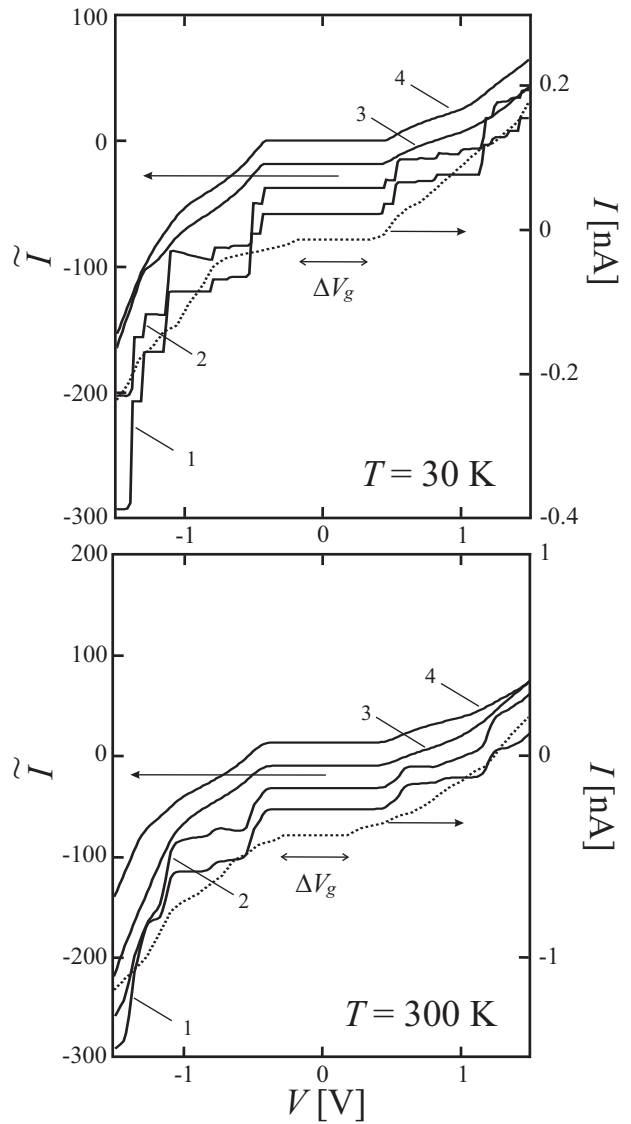}
\caption{Calculated $\tilde{I}-V$  curves for structure based on
spherical cluster of the radius of 1 nm at two values of two
temperature of structure and \emph{1} -- $T_{e}=T_{i}=T$ (i.e.
without the account of broadening and overheating), \emph{2} -- with
the account of only overheating, \emph{3} -- with the account of
only broadening, \emph{4} -- with by the account of both the
broadening and overheating. Experimental curve \cite{Ohgi,Ohgi-2003}
is given by dotted line. For presentation purposes, the curves are
shifted slightly in a vertical direction. In experiments, the
current gap width $\Delta V_{g}$ is $0.55\pm 0.1$ eV at $T=30$Ê
\cite{Ohgi-2001} and 300 K \cite{Ohgi-2003}.} \label{BP-2009-Fig4}
\end{figure}
\begin{figure}[!t!b!p]
\centering
\includegraphics [width=.45\textwidth] {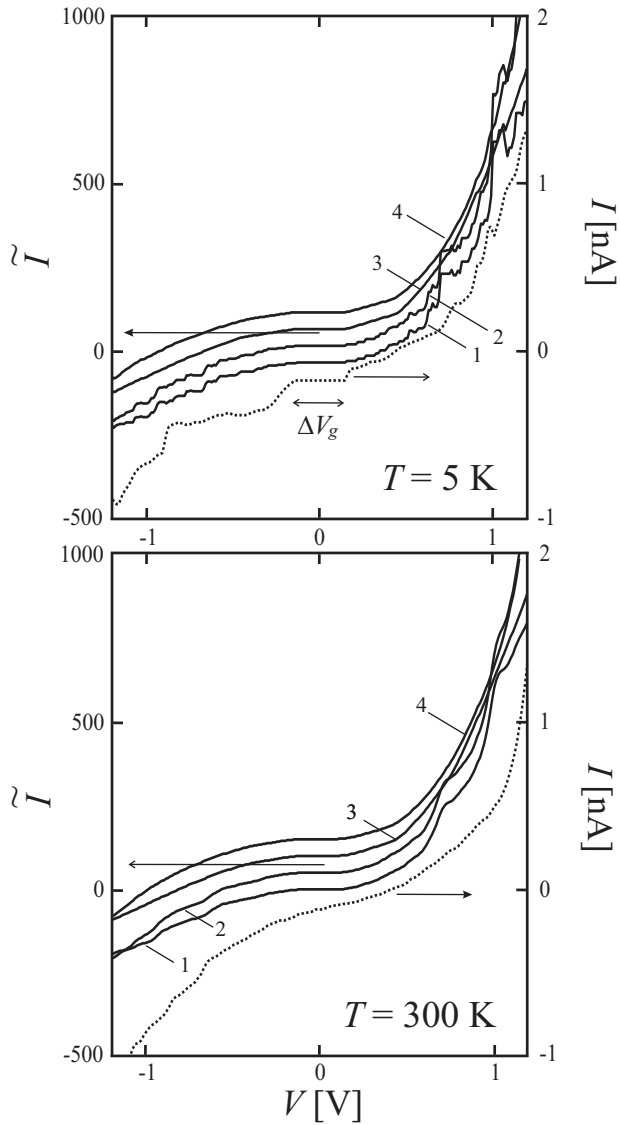}
\caption{Calculated $\tilde{I}-V$  curves for structure based on
cluster-disc of the radius 2 nm. Experimental curve is taken from
Ref. \cite{Wang}:  $\Delta V_{g}=0.3\pm 0.075$ eV at $T=5$K and
$\Delta V_{g}\rightarrow 0$  at $T=300$K.} \label{BP-2009-Fig5}
\end{figure}

The broadening of the levels  mimics a quasi-continuous spectrum in
a cluster. The calculation of broadening is performed  for disk in
absence bias of a voltage. This approximation has to be considered
as an estimate of a minimum broadening for the whole $I(V)$ curve.
In the disc-shaped dot, the electron states are realized only with
``subbands'' $n_{z}=1$ and 2. States with $n_{z}=1$ almost do not
decay. The broadening of the levels with $n_{z}=2$ is approximately
0.7 eV. Estimation of broadening in spheres is rather rough, since
it does not take into account obvious three-dimensionality of the
problem.

Width $\Delta V_{g}$ for non-magic clusters is determined only by
charging energy $\widetilde{E}_{\rm C}$. An overheating in current
gap is not substantial, because feeding energy is minimal. At
$T=300$ K our calculation gives $\Delta V_{g}$, in satisfactory
agreement with experiment for structure based on the disc. The
result for a disk however is quite sensitive to the temperature. It
can be due to the change of cluster shape and its metallic
properties because of the thermal fluctuations.

Steps of the Coulomb and quantum staircases are clearly visible
outside the region of current gap in $I-V$ curves, which are
calculated neglecting the broadening and the overheating,. The
broadening and the overheating give similar results in smoothing of
staircases for granule-molecule at low temperatures. However, a role
of the broadening is much more significant.

\section{Conclusion}

In this work the semi-empiric estimations of two mechanisms are
performed: (i)  broadening of electronic levels due to a tunnel
effect, (ii) heating of electronic gas in the isolated metal
clusters in presence of bias voltage. The calculations are carried
out for two gold clusters of close volume and different shapes
(cylindrical and spherical).

A calculation scheme of $S-$matrix poles for the broadening effect
in simplest model of rectangular barriers of three-electrode
structure is suggested. For monoatomic disc, containing
approximately 250 atoms, the broadening of ``work subband'' $\gamma$
is 0.7 eV, provided that the difference between discrete levels is
close to 0.2 eV in the vicinity of the granule Fermi level
$\Delta\varepsilon_{\rm F}$.

In the framework of two-temperature model of metal cluster, and by
using a size dependence of the Debye frequency, the effective
electron temperature vs. bias voltage is found approximately. The
strong dependence of electronic kinetic temperature vs. voltage is
found. For helium temperature of ion subsystem, the heating
temperature of electrons in a quantum disk is almost one order of
magnitude higher than that in a sphere; it achieves thousands of
Kelvins.

We suggest an explanation for the effect of strong smoothing of
current-voltage curves in structures based on clusters consisting of
accountable number of atoms which was observed experimentally at low
temperatures. We believe that this effect can be attributed to the
level broadening and electron subsystem overheating, with the
influence of broadening being much more important.

The indicated mechanisms can violate the basic inequalities, which
have to be fulfilled for single-electronic devices to be able to
work
$$
\widetilde{E}_{\rm C},\,\Delta\varepsilon_{\rm F}\gg k_{\rm B}T,
$$
because, in these conditions, it is necessary to replace
$\Delta\varepsilon_{\rm F}$ by $\Delta\varepsilon_{\rm
F}-\gamma_{\rm F}$ and $T$ by $T_{e}$. ($\gamma_{\rm F}$ is the
``average'' broadening of discrete levels in the vicinity of the
granule's Fermi level).

Fabrication of stable in sizes and shapes elements is one of the key
problems of nanoelectronics. Structures created on metallic clusters
is not while succeeded. This problem, possibly, can be realized on
the clusters by sort Zn$@$C$_{28}$ \cite{Ivano} They convenient to
those that tunneling transitions can be exactly enough organized as
dielectric shell round the atom of metal steady.  It turns out to be
a difficult task to describe a device based on such a cluster by
using simple models: in particular, because it is not possible to
use a charging energy $\widetilde{E}_{\rm C}$ as an informative
parameter. For this purpose it is necessary to know an electron
affinity and ionization potential of metal atom in the shell of
carbon atoms. Moreover effect of overheating will be absent owing to
lack of a electron gas.

A thermoemission current depends exponentially on the ratio of
granule electron work function and the kinetic temperature. Change
of current is substantial, provided that the temperature is changed
in tens of times. We suppose that the broadening of the levels must
be also taken into account for the proper description of thermo- and
photoemission in similar structures.

We are grateful to W. V. Pogosov for reading the manuscript. This
work was supported by the Ministry of Education and Science of the
Ukraine.


\begin{references}


\thispagestyle{empty}

\bibitem{Gub-98}E.S. Soldatov, V.V. Khanin, A.S. Trifonov, S.P. Gubin, V.V. Kolesov,
D.E. Presnov, S.A. Yakovenko, G.V. Khomutov, and A.N. Korotkov.
Uspekhi Fiz. Nauk \textbf{168}, 217 (1998)  [Physics-Uspekhi
\textbf{41} 202 (1998)].

\bibitem{Ralph}J. von Delft, and D.C. Ralph. Phys. Rep.
\textbf{345}, 61 (2001).

\bibitem{Wang}B. Wang, X. Xiao, X. Huang,  P. Sheng, J.G. Hou.
Appl. Phys. Lett. \textbf{77},  1179 (2000).

\bibitem{Hou}J.G. Hou,  B. Wang, J. Yang, X. R. Wang, H.Q. Wang,
Q. Zhu, X. Xiao. Phys. Rev. Lett. \textbf{86},  5321 (2001).

\bibitem{Ohgi}T. Ohgi, D. Fujita.  Physica E.  \textbf{18},  349 (2003).

\bibitem{Ohgi-2003}T. Ohgi,  D. Fujita. Surf. Sci. \textbf{532-535}, 294 (2003).

\bibitem{FTT-2006}V.V. Pogosov and E.V. Vasyutin. Nanotechnology. \textbf{17}, 3366
(2006).

\bibitem{LetJTF-2007}V.V. Pogosov, E.V. Vasyutin, and A.V. Babich.  Pisma v
Zh. Tekhn. Fiz. \textbf{33}, 1 (2007) [Techn. Phys. Lett.
\textbf{33} 719 (2007); arXiv:cond-mat/0611551].

\bibitem{ginz}V.L. Ginzburg and V.P. Shabanskii, Dokl. Akad. Nauk SSSR
\textbf{100},  445 (1955).

\bibitem{kag}M.I. Kaganov, I.M. Lifshitz, and L.V. Tanatarov, Zh. Eksp. Teor.
Fiz. \textbf{31}, 232 (1956)[Sov. Phys. JETP \textbf{4}, 173
(1957)].

\bibitem{zl}Z. Lin, L.V. Zhigilei, V. Celli.    Phys. Rev. B  \textbf{77},
 075133 (2008).

\bibitem{Yan}Y.-F. Zhang, J.-F. Jia, T.-Z. Han, Z. Tang, Q.-T. Shen, Y. Guo, Z.Q. Qiu,  Q.-K. Xue.
Phys. Rev. Lett. \textbf{95},  096802 (2005).

\bibitem{Gloskovskii}A.
Gloskovskii, D.A. Valdaitsev, M. Cinchetti, S.A. Nepijko, J. Lange,
M. Aeschlimann, M. Bauer, M. Klimenkov, L.V. Viduta, P.M. Tomchuk,
and G. Sch\"{o}nhense, Phys. Rev. B \textbf{77}, 195427 (2008).

\bibitem{Kopidakis}G. Kopidakis, C. M. Soukoulis, E. N.
Economou.  Phys. Rev. B  \textbf{49},  7036 (1994).

\bibitem{Qu}S.-X. Qu, A.N. Cleland, M.R. Geller.  Phys. Rev. B \textbf{72},  224301
(2005).

\bibitem{Vega}L. de la Vega, A. Mart\'{e}n-Rodero, N. Agra\"{\i}t, A. Levy Yeyati.
 Phys. Rev. B  \textbf{73},   075428 (2006).

\bibitem{Agosta}R. D'Agosta, Na Sai, M. Di Ventra.  Nano Lett. \textbf{6}, 2935
(2006).

\bibitem{Galperin}M. Galperin, M.A. Ratner, A. Nitzan. J. Phys.: Cond. Matt.
\textbf{19},  103201 (2007).

\bibitem{BLT}E.D. Belotski, S.P. Luk'yanets, P.M. Tomchuk. Zh. Eksp.
Teor. Fiz. \textbf{101}, 163 (1992) [Sov. Phys. JETP \textbf{74}
(1992) 88].

\bibitem{Fed}R.D. Fedorovich, A.G. Naumovets, P.M. Tomchuk.
Phys. Rep. \textbf{328} 73 (2000).

\bibitem{Bil}Y. Bilotsky, P.M. Tomchuk.   Surf. Sci. \textbf{602},
 383 (2008).

\bibitem{Singh}N. Singh. arXiv:cond-mat/0702331.

\bibitem{Ovadia}M. Ovadia, B. Sacepe, and D. Shahar. Phys. Rev. Lett. \textbf{102}, 176802
(2009).

\bibitem{Kres}V.V. Kresin, Yu. N. Ovchinnikov.  Phys. Rev. B
\textbf{73}, 115412 (2006).

\bibitem{Pushpa}R. Pushpa, U. Waghmare,  S. Narasimhan.
Phys. Rev. B \textbf{77},  045427 (2008).

\bibitem{Rethfeld}B. Rethfeld, A. Kaiser, M. Vicanek, G. Simon.  Phys. Rev. B  \textbf{65},  214303
(2002).

\bibitem{Rethfeld}B. Rethfeld, A. Kaiser, M. Vicanek, G. Simon.  Phys. Rev. B  \textbf{65},  214303
(2002).

\bibitem{anis-2004}B. Rethfeld, K.  Sokolowski-Tinten,  D.
von der Linde, S.I. Anisimov. Appl. Phys. A  \textbf{79},  767
(2004).

\bibitem{Ji}L. Jiang, H.-L. Tsai.  J. Heat Transfer \textbf{127},  1167 (2005).

\bibitem{Gantefor}G. Gantefor, W.  Eberhardt, H.  Weidele, D.  Kreisle, E.  Recknagel.
Phys. Rev. Lett. \textbf{77},   4524  (1996).

\bibitem{Maier}M. Maier, M. Schatze, G. Wrigge, M.A. Hoffmann, P. Didier,
B.V. Issendorff. Int. J. Mass Spectr. \textbf{252}, 157 (2006).

\bibitem{Co}P.R. Couchman, F.E.  Karasz.   Phys. Lett. A
\textbf{62},   59 (1977).

\bibitem{Balerna}A. Balerna, S. Mobilio. Phys. Rev. B
\textbf{34},  2293 (1986).

\bibitem{Gu}M.X. Gu, C.Q. Sun, Z. Chen, T.C. Au Yeung, S. Li,
C. M. Tan, V. Nosik. Phys. Rev. B \textbf{75},
 125403  (2007).

\bibitem{Zhang}C.-H. Zhang, F. Kassubek, C. A. Stafford. Phys. Rev. B  \textbf{68},
 165414 (2003).

\bibitem{Urban}D.F. Urban, C.A. Stafford, H. Grabert. Phys. Rev. B
\textbf{75}, 205428 (2007).

\bibitem{Mar}A. Maradudin, R. Wallis. Phys. Rev. \textbf{148}, 945 (1966).

\bibitem{288}A. I. Baz', Ya. B. Zel'dovich, and A. M. Perelomov, Scattering,
Reactions and Decays in Nonrelativistic Quantum Mechanics, transl.
of 1st Russ. ed. (Nauka, Moscow, 1971, 2nd ed.; Israel Program for
Scientific Translations, Jerusalem, 1966).

\bibitem{PHR-2005}V. V. Pogosov, V.  P. Kurbatsky,  and E.
V. Vasyutin.    Phys. Rev. B \textbf{71}, 195410 (2005).

\bibitem{Ohgi-2001}T. Ohgi, H.-Y. Sheng, Z.-C. Dong, H. Nejoh,
D. Fujita  // Appl. Phys. Lett.  \textbf{79},  2453 (2001).

\bibitem{Imamura}H. Imamura, J. Chiba, S. Mitani, K. Takanashi, S. Takahashi, S.
Maekawa, H. Fujimori. Phys. Rev. B \textbf{61}, 46 (2000).

\bibitem{Ivano}M.J.S. Dewar, E.G. Zoebisch, E.F. Healy. J. Am. Chem. Soc.
\textbf{107}, 3902 (1985).


\end{references}
\end{document}